\documentclass[fleqn,usenatbib]{mnras}

\usepackage{newtxtext,newtxmath}

\usepackage[T1]{fontenc}

\let\VANthebibliography\thebibliography
\def\thebibliography{\DeclareRobustCommand{\VAN}[3]{##3}\VANthebibliography}

\usepackage{makecell}
\usepackage{amsmath}
\usepackage{booktabs}
\usepackage{lineno}
\usepackage{threeparttable}
\usepackage{makecell}
\usepackage[squaren]{SIunits}
\usepackage{comment}
\usepackage{xcolor}
\usepackage{multirow}
\usepackage{appendix}
\usepackage{longtable,booktabs}
\usepackage{float}
\usepackage{graphicx}	
\usepackage{amsmath}	
\usepackage{amssymb}	
\usepackage[squaren]{SIunits}
\usepackage[export]{adjustbox}

\title[Spectral lag of SGR J1935+2154]{Discovery of the linear energy-dependence of the spectral lag of X-ray bursts from SGR J1935+2154}

\author[S. Xiao et al.]{
Shuo Xiao,$^{1,2}$
You-Li Tuo,$^{3,4}$\thanks{E-mail: tuo@astro.uni-tuebingen.de}
Shuang-Nan Zhang,$^{4,5}$\thanks{E-mail: zhangsn@ihep.ac.cn}
Shao-Lin Xiong,$^{4}$\thanks{E-mail: xiongsl@ihep.ac.cn}
Lin Lin,$^{6}$
Yan-Qiu Zhang,$^{4,5}$
\newauthor
Yue Wang,$^{4,5}$
Wang-Chen Xue,$^{4,5}$
Ce Cai,$^{7}$
He Gao,$^{6}$
Cheng-Kui Li,$^{4}$
Xiao-Bo Li,$^{4}$
Chao Zheng$^{4,5}$,  
\newauthor
Jia-Cong Liu$^{4,5}$  
Ping Wang,$^{4}$
Jin Wang,$^{4}$
Wen-Xi Peng,$^{4}$
Cong-Zhan Liu,$^{4}$
Xin-Qiao Li,$^{4}$
Xiang-Yang Wen,$^{4}$
\newauthor
Zheng-Hua An,$^{4}$
Li-Ming Song,$^{4}$
Shi-Jie Zheng$^{4}$, 
Fan Zhang$^{4}$,
Ai-Jun Dong,$^{1,2}$ 
Wei Xie,$^{1,2}$
Jian-Chao Feng,$^{1,2}$
\newauthor
Qing-Bo Ma,$^{1,2}$
De-Hua Wang,$^{1,2}$
Xi-Hong Luo,$^{1,2}$
Shi-Jun Dang,$^{1,2}$
Lun-Hua Shang,$^{1,2}$
Qi-Jun Zhi,$^{1,2}$
Ti-Pei Li$^{4,8}$
\\
$^{1}$Guizhou Provincial Key Laboratory of Radio Astronomy and Data Processing, Guizhou Normal University, Guiyang 550001, People’s Republic of China\\
$^{2}$School of Physics and Electronic Science, Guizhou Normal University, Guiyang 550001, People’s Republic of China\\
$^{3}$Institut für Astronomie und Astrophysik, University of Tübingen, Sand 1, 72076 Tübingen, Germany\\
$^{4}$Key Laboratory of Particle Astrophysics, Institute of High Energy Physics, Chinese Academy of Sciences, Beijing 100049, China\\
$^{5}$University of Chinese Academy of Sciences, Chinese Academy of Sciences, Beijing 100049, China\\
$^{6}$Department of Astronomy, Beijing Normal University, Beijing 100875, People’s Republic of China\\
$^{7}$College of Physics, Hebei Normal University, 20 South Erhuan Road, Shijiazhuang, 050024, China\\
$^{8}$Department of Astronomy, Tsinghua University, Beijing 100084, People’s Republic of China\\
}

\date{Accepted XXX. Received YYY; in original form ZZZ}

\pubyear{2022}

\begin{document}
\label{firstpage}
\pagerange{\pageref{firstpage}--\pageref{lastpage}}
\maketitle

\begin{abstract}
Spectral lag of the low-energy photons with respect to the high-energy ones is a common astrophysical phenomenon (such as Gamma-ray bursts and the Crab pulsar) and may serve as a key probe to the underlying radiation mechanism. However, spectral lag in keV range of the magnetar bursts has not been systematically studied yet. In this work, we perform a detailed spectral lag analysis with the Li–CCF method for SGR J1935+2154 bursts observed by {\it Insight}-HXMT, GECAM and Fermi/GBM from July 2014 to Jan 2022. We discover that the spectral lags of about 61\% (non-zero significance >1$\sigma$) bursts from SGR J1935+2154 are linearly dependent on the photon energy ($E$) with $t_{\rm lag}(E)=\alpha (E/{\rm keV})+C$, which may be explained by a linear change of the temperature of the blackbody-emitting plasma with time. The distribution of the slope ($\alpha$) approximately follows a Gaussian function with mean and standard deviation of 0.02 ms/keV (i.e. high-energy photons arrive earlier) and 0.02 ms/keV, respectively. We also find that the distribution can be well fitted with three Gaussians with mean values of $\sim$ -0.009, 0.013 and 0.039 ms/keV, which may correspond to different origins of the bursts.
These spectral lag features may have important implications on the magnetar bursts.

\end{abstract}

\begin{keywords}
stars: magnetars
\end{keywords}


\section{Introduction}
Spectral lag of the low-energy photons with respect to the high-energy ones is a common phenomenon in astronomical sources. It is conventionally defined as positive lag when low-energy photons follow high-energy photons. For example, Gamma-ray bursts (GRBs) have positive or negative lag which usually have an energy dependence of $t_{\rm lag}(E)=A \ln (E/{\rm keV})+C$ for short GRBs (SGRBs) \citep{minaev2014catalog, pozanenko2018grb, xiao2022quasi} and $t_{\rm lag}(E)=C-A(E/{\rm keV})^{-\beta}$ for long GRBs (LGRBs) with positive lags \citep{shao2017new,xiao2022energetic}, which may be explained as “curvature” effect \citep{zhang2009discerning}. Instead, Crab pulsar have a negative lag with $t_{\rm lag}(E)=-A(E/{\rm keV})+C$ ($A\approx 0.27\ {\rm \mu s/keV}$) \citep{molkov2009absolute,tuo2022orbit}, which can be modeled with two independent components with different spectral and phase distribution  \citep{massaro2000fine}. 

On the other hand, the spectral lag of magnetar bursts is usually regarded as negligible in previous studies, such as GRB 200415A with magnetar origin \citep{yang2020grb}. Recently, a Galactic fast radio burst (FRB) at 14:34:24 UTC on April 28, 2020 is associated with an X-ray burst (XRB) from the Galactic magnetar SGR J1935+2154 \citep{israel2016discovery} indicating that at least some FRBs originate from magnetar activities (e.g. \citealp{li2021hxmt,bochenek2020fast,mereghetti2020integral,younes2021broadband,ridnaia2021peculiar,tavani2021x}). Interestingly, two narrow X-ray peaks are found to be closely associated with the two radio pulses \citep{li2021hxmt}. It was also reported that the X-ray peak has a delay of $\sim$ 6.5 ms with respect to the 1.4 GHz radio pulse \citep{mereghetti2020integral}. Although some models can explain this feature (e.g. \citealp{dai2020magnetar}), the radiation mechanism is not very clear, in particular, why only this XRB, out of many, has been observed to be associated with FRB \citep{lin2020no}?

SGR J1935+2154 is one of the most active magnetars since its discovery in 2014 \citep{israel2016discovery} and its bursts may exhibit a periodic behavior \citep{xie2022revisit, zou2021periodicity}. From July 2014 to Jan 2022, there are about 700 bursts observed by {\it Insight}-HXMT \citep{li2021hxmt,cai2022insight}, GECAM \citep{xiao2022energetic,xie2022revisit} and Fermi/GBM \citep{lin2020fermi,zou2021periodicity}. These three satellites have high temporal resolution of $2\ {\rm \mu s}$ \citep{liu2020high,xiao2020deadtime}, $0.1\ {\rm \mu s}$ \citep{xiao2022ground} and $2\ {\rm \mu s}$ \citep{meegan2009fermi}, as well as wide energy range of 0.2 keV - 3 MeV \citep{zhang2020overview}, 10 keV - 8 MeV \citep{li2021inflight} and 8 keV - 40 MeV \citep{meegan2009fermi}, respectively, which are beneficial for spectral and timing analyses. In particular, spectral lags have not been studied so far on these bursts observed by these three satellites, which may serve as key clues to the underlying radiation mechanism of SGR J1935+2154 bursts, and may be used as a probe of spectral evolution whenever a detailed time-resolved spectral analysis is not possible.


In this paper, we collect a nearly complete sample of bursts from SGR J1935+2154 observed by {\it Insight}-HXMT, GECAM and Fermi/GBM from July 2014 to Jan 2022 (Section 2.1), and used the Li-CCF method (Section 2.2) to calculate the spectral lag. The results are analyzed and reported in Section 3. Finally, discussion and summary of the results are given.

\begin{figure*}
\centering
\begin{minipage}[t]{0.46\textwidth}
\centering
\includegraphics[width=\columnwidth]{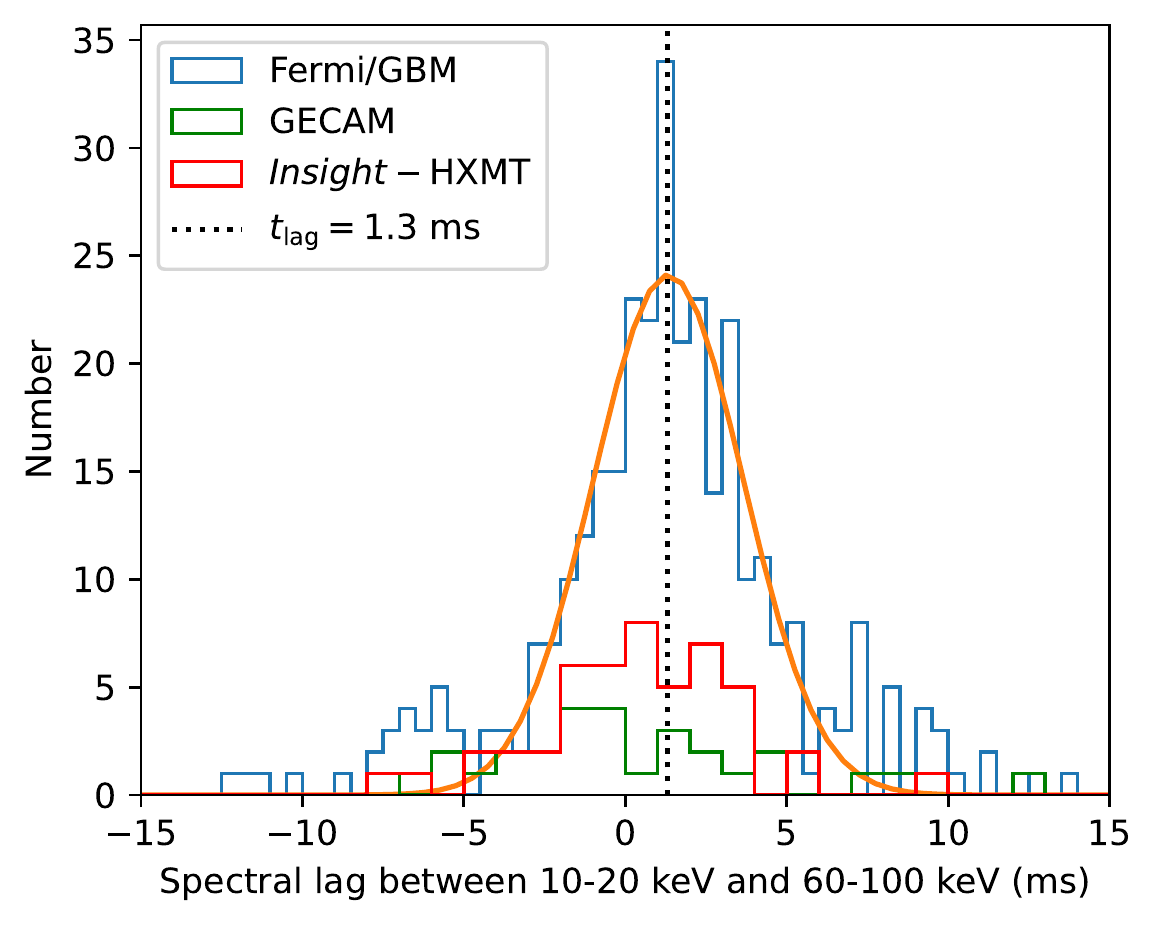}
\end{minipage}
\begin{minipage}[t]{0.46\textwidth}
\centering
\includegraphics[width=\columnwidth]{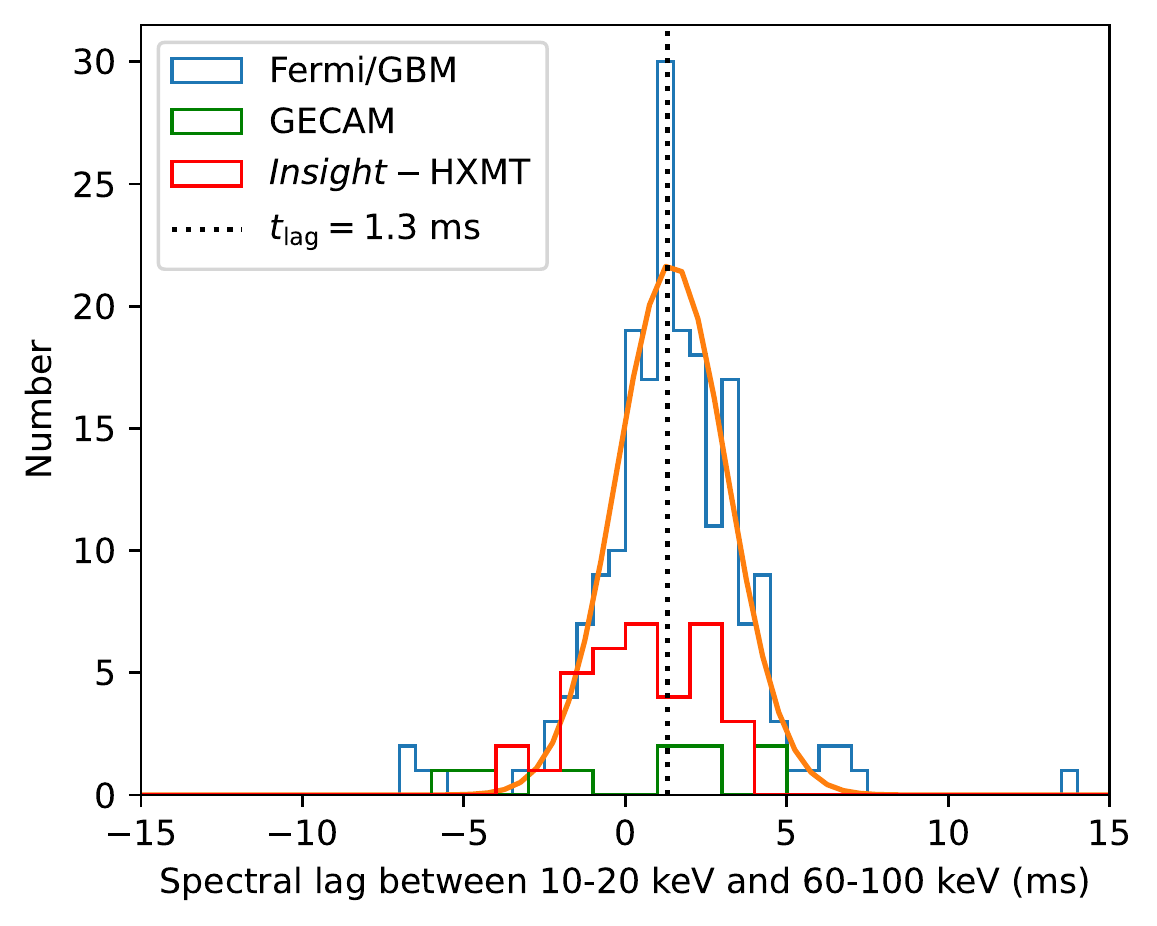}
\end{minipage}
\caption{The left panel shows the distribution of the spectral lags (with error$<$20 ms) between 10-20 keV and 60-100 keV for each burst observed by {\it Insight}-HXMT or GECAM or Fermi/GBM. The right panel shows the distribution of the spectral lags with error$<$5 ms. The fitted yellow lines are Gaussian functions with mean and standard deviation of 1.3 ms and 2.3 ms for the left panel, 1.4 ms and 1.7 ms for the right panel. Note that most of the samples from the three satellites are different.}\label{lag_2}
\end{figure*}

\begin{figure}
\centering
\centering
\includegraphics[width=\columnwidth]{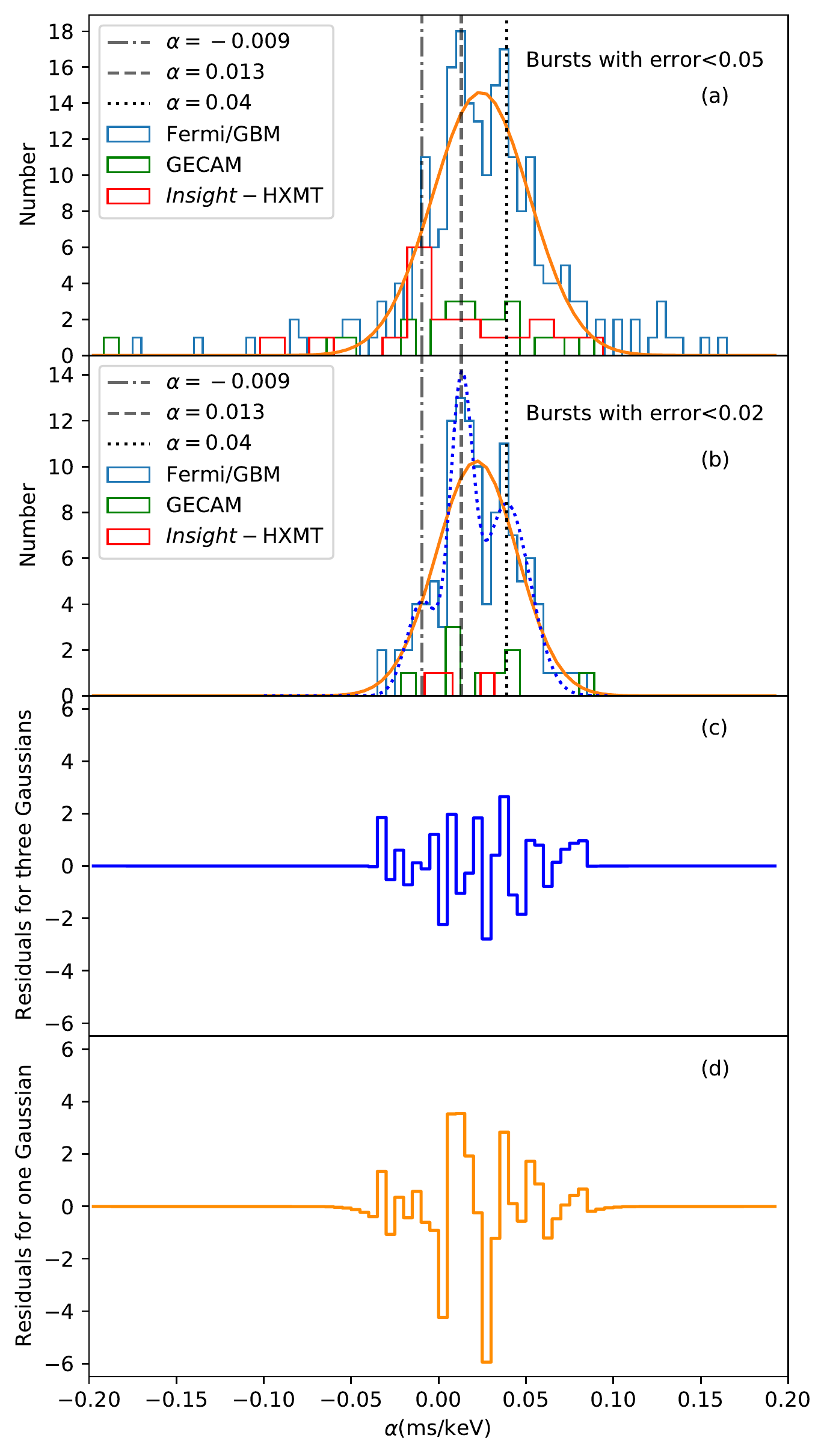}
\caption{Panel (a): distributions of the slopes (with the error of $\alpha$ $<$0.05 ms/keV) of the energy-spectral lag relation $t_{\rm lag}(E) =\alpha (E/{\rm keV})+C$ for each burst observed by {\it Insight}-HXMT or GECAM or Fermi/GBM, respectively. The fitted yellow line is fitted by one Gaussian function with mean and standard deviation of 0.02 and 0.04 ms/keV.
Panel (b): distributions of the slopes with the error of $\alpha$ $<$0.02 ms/keV. The slopes are centered around the three peaks, -0.009, 0.013 and 0.039 ms/keV. The distributions 
obtained by GBM and GECAM are consistent, whereas that obtained by HXMT is more concentrated around -0.009 ms/keV. The fitted yellow line is one Gaussian function with mean and standard deviation of 0.02 and 0.02 ms/keV. The blue dashed lines are fitted by three Gaussians simultaneously. 
Panels (c) and (d): the residuals for three Gaussians and one Gaussian models, respectively. 
Note that most of the samples from the three satellites are different.}\label{alphas}
\end{figure}

\begin{figure*}
\centering
\begin{minipage}[t]{0.46\textwidth}
\centering
\includegraphics[width=\columnwidth]{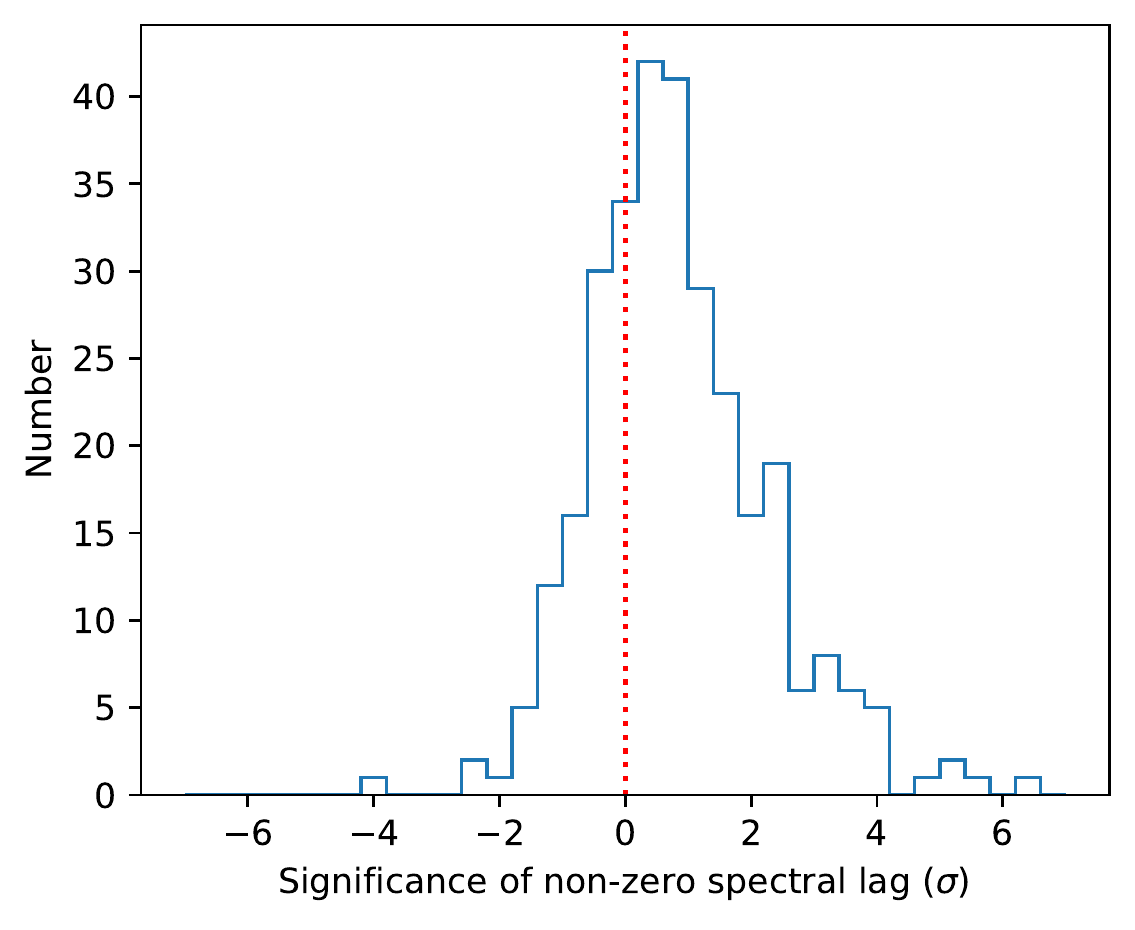}
\end{minipage}
\begin{minipage}[t]{0.46\textwidth}
\centering
\includegraphics[width=\columnwidth]{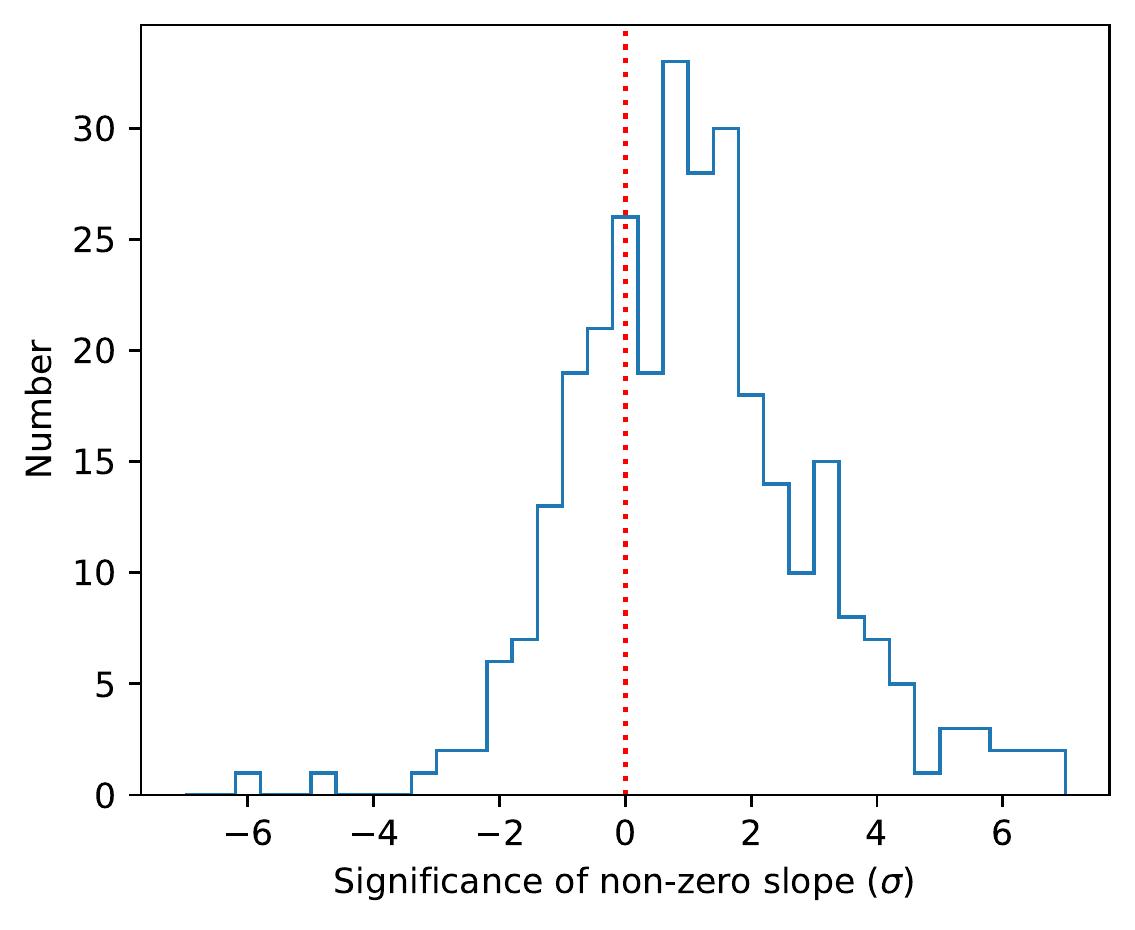}
\end{minipage}
\caption{The left panel shows the distribution of the significance of non-zero spectral lag between 10-20 keV and 60-100 keV.
The right panel shows the distribution of the significance of non-zero slope of the energy-spectral lag relation $t_{\rm lag}(E) =\alpha (E/{\rm keV})+C$.}\label{sigmas}
\end{figure*}

\begin{figure*}
\centering
\begin{minipage}[t]{0.46\textwidth}
\centering
\includegraphics[width=\columnwidth]{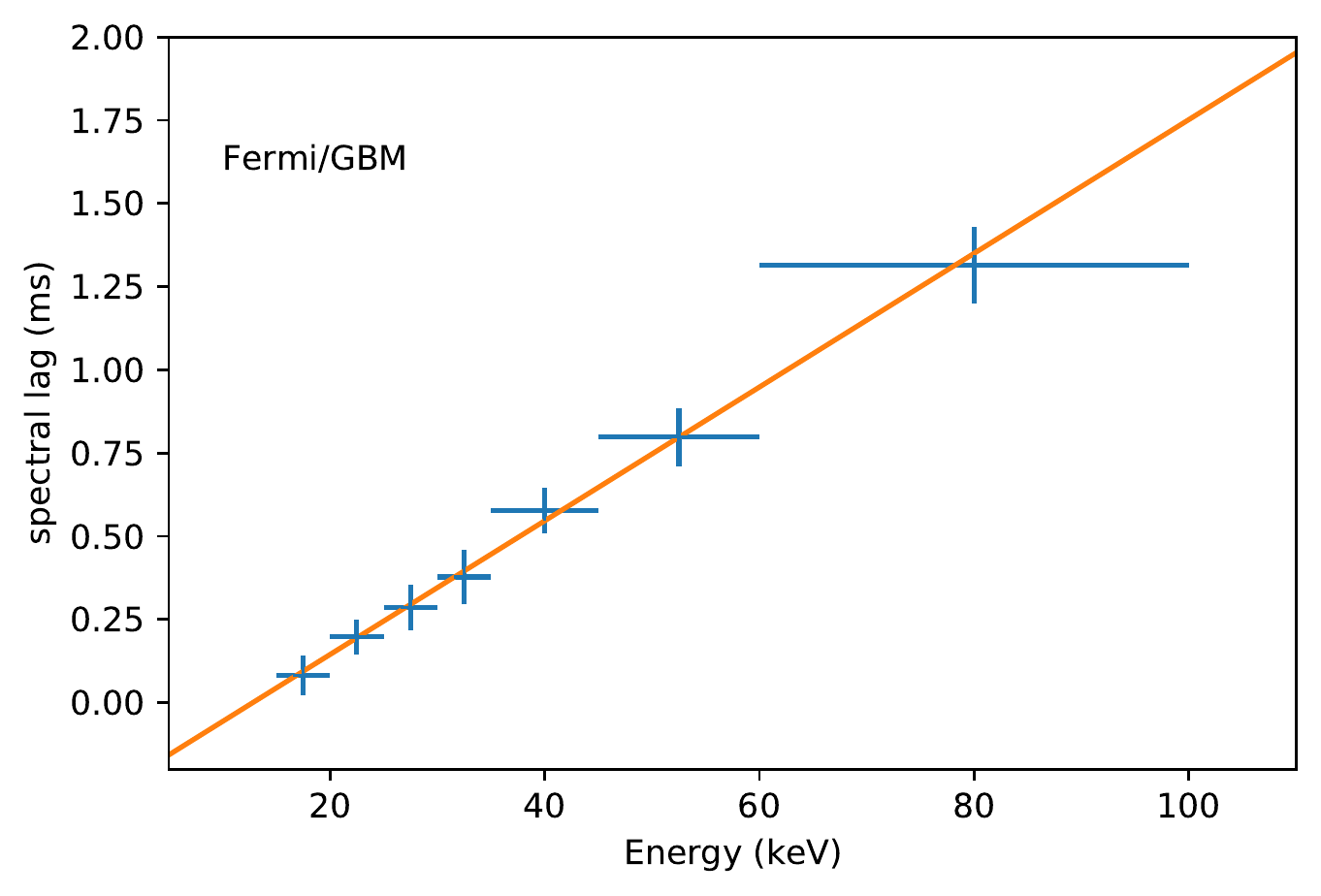}
\end{minipage}
\begin{minipage}[t]{0.46\textwidth}
\centering
\includegraphics[width=\columnwidth]{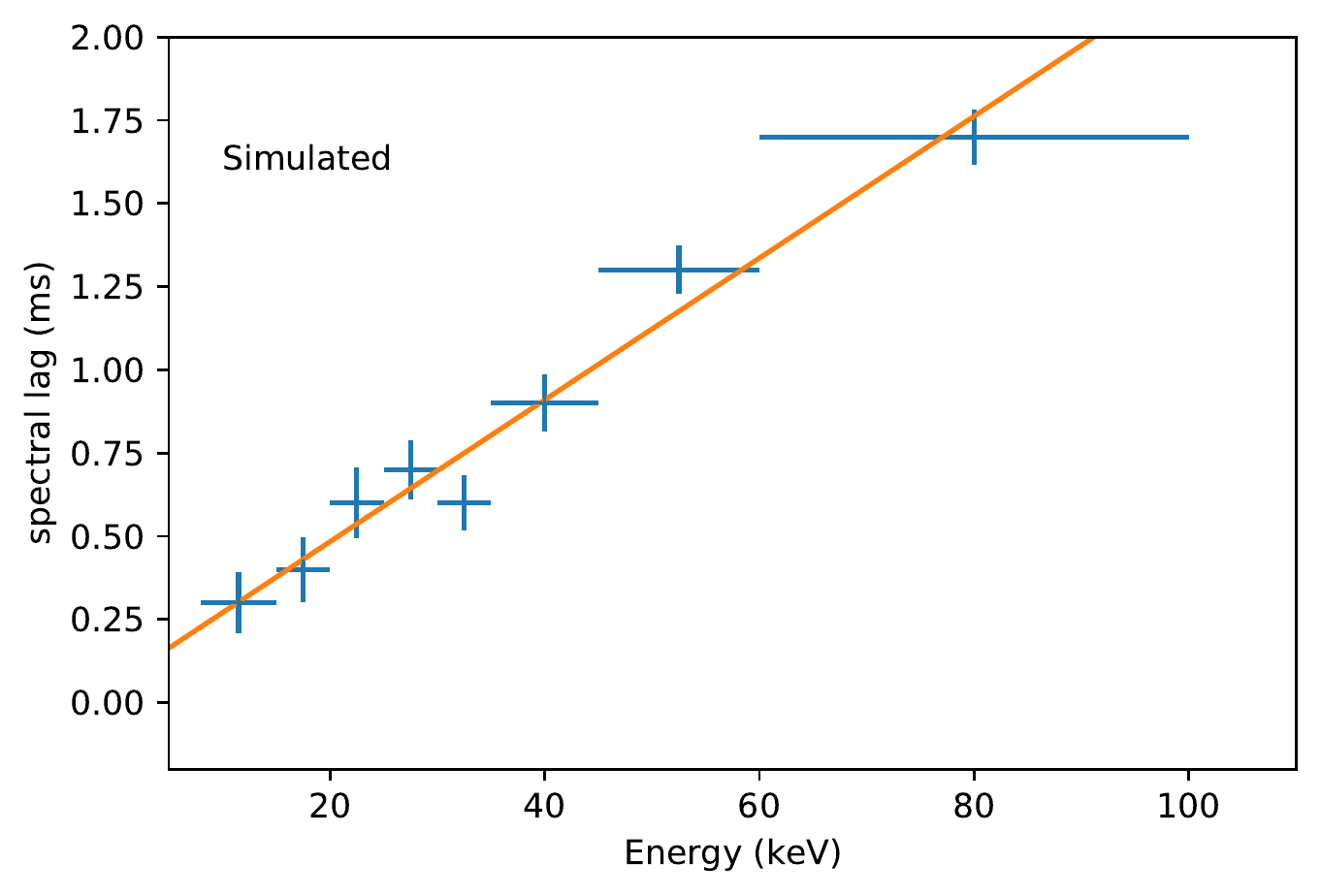}
\end{minipage}
\caption{The left panel shows spectral lags compared to 8-15 keV of joint fitting (see Fig. A1) of bursts from SGR J1935+2154 observed by GBM, the orange line is fitted with $t_{\rm lag}(E)=0.020^{+0.001}_{-0.001}(E/{\rm keV})-0.258$ ms.
The right panel shows the spectral lags with energy for the simulated linear decrease of temperature with time for a blackbody ($kT=-12\ (t/{\rm s})+14$ keV) in a lightcurve, the orange line is fitted with $t_{\rm lag}(E)\approx0.02 (E/{\rm keV})+C$, which is consistent with the bursts from SGR J1935+2154. }\label{lag_fit_joint}
\end{figure*}

\begin{figure*}
\centering
\begin{minipage}[t]{0.49\textwidth}
\centering
\includegraphics[width=\columnwidth]{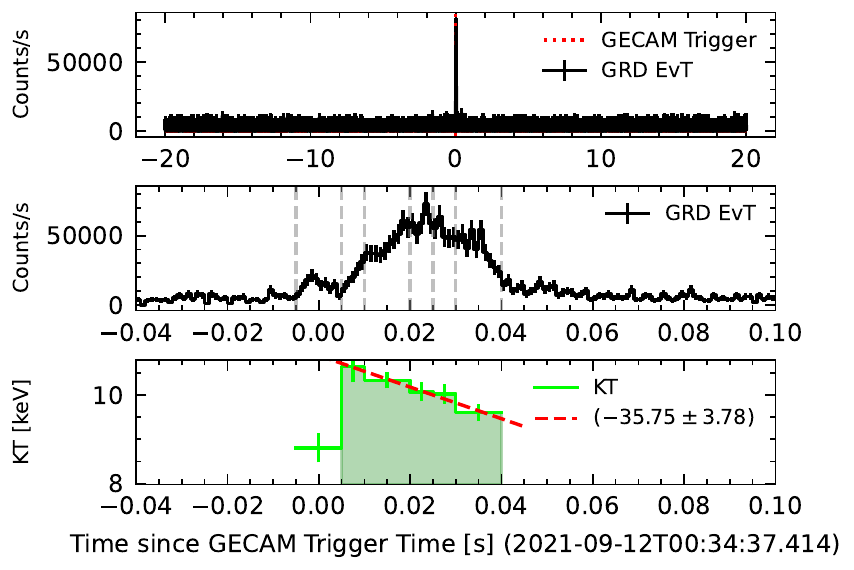}
\end{minipage}
\begin{minipage}[t]{0.49\textwidth}
\centering
\includegraphics[width=\columnwidth]{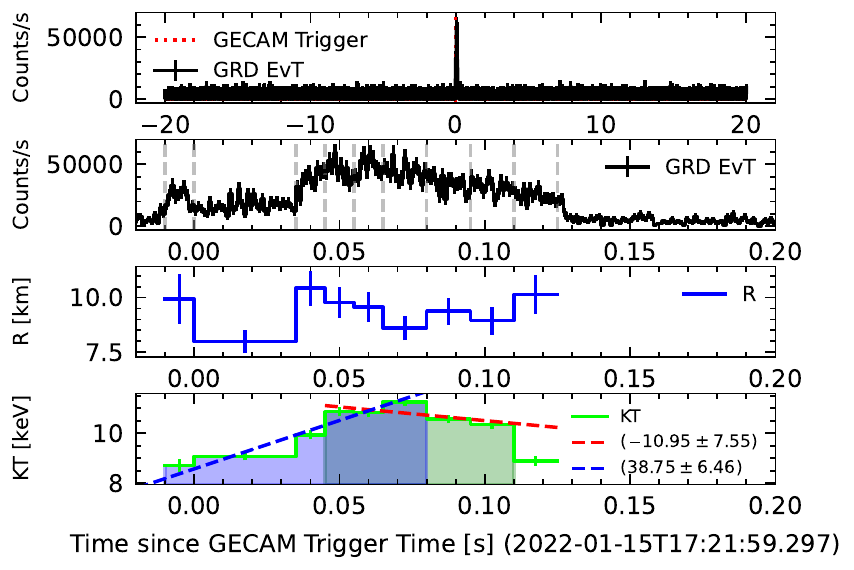}
\end{minipage}
\caption{The light curves and the evolution of the Blackbody temperature (KT) with time. The red and blue dashed lines are obtained by fitting with $KT=at+C$, respectively, and the numbers in parentheses are the result of slope $a$. The left and right panels are bursts from SGR J1935+2154 at UTC 2021-09-12T00:34:37.414 and 2022-01-15T17:21:59.297, respectively.}\label{spec_fit}
\end{figure*}

\begin{table*}
\caption{\centering Parameters for fitting the distribution of slope $\alpha$ using one and three Gaussian models, respectively.}\label{fit_table}
\begin{tabular*}{\hsize}{@{}@{\extracolsep{\fill}}cccc@{}}
	\hline
 Fitting model & \makecell[c]{Amplitude} & \makecell[c]{Mean (ms/keV)} & \makecell[c]{Standard deviation}\\
	\hline
 One Gaussian & \makecell[c]{10.3$\pm$0.95} &\makecell[c]{0.022$\pm$0.003} & \makecell[c]{0.023$\pm$0.002} \\
Three Gaussians &\makecell[c]{4.1$\pm$2.2 \\12.7$\pm$2.5 \\8.4$\pm$2.3} 
 & \makecell[c]{-0.009$\pm$0.005 \\0.013$\pm$0.002 \\ 0.039$\pm 0.004$}
 & \makecell[c]{0.009$\pm$0.005 \\0.006$\pm$0.002 \\ 0.013$\pm$0.005} \\
	\hline
\end{tabular*}\\
\end{table*}

\section{SAMPLE SELECTION and Methodology}
\subsection{Data selection and Preparation}
The sample of bursts from SGR J1935+2154 observed by {\it Insight}-HXMT, GECAM and Fermi/GBM from July 2014 to Jan 2022 is collected from literature \citep{lin2020fermi,li2021hxmt,zou2021periodicity,xiao2022energetic,cai2022insight,xie2022revisit}.

For bursts observed by {\it Insight}-HXMT, the data of the three telescopes with different energy ranges are used, including the High Energy X-ray telescope (HE) with 20-250 keV \citep{liu2020high}, the Medium Energy X-ray
telescope (ME) with 5-30 keV \citep{cao2020medium} and the Low Energy X-ray telescope
(LE) with 1-15 keV \citep{chen2020low}. We select 1-10, 10-30, 30-100 keV for LE, ME and HE according to the effective area \citep{zhang2020overview}, respectively.
 We filter out these bursts that saturated the HXMT detectors due to their extreme brightness \citep{xiao2020deadtime}. Besides, since the dead time of LE is negligible \citep{chen2020low} and the HE dead time for most bursts are below 10\% \citep{xiao2020deadtime}, dead time correction is not required for LE and HE. However, dead time correction is required for ME because it can reach up to 50\% (e.g. \citealp{li2021hxmt}). Finally, it is worth noting that the time system of LE needs to be corrected for the effect of a delay of about 860 $\mu s$ compared to ME and HE \citep{tuo2022orbit}. After this correction, the time accuracy of the three telescopes of HXMT is within 30 $\mu$s \citep{tuo2022orbit}.

For bursts from SGR J1935+2154 observed by GECAM and GBM, the data in energy ranges of 15-100 keV and 8-100 keV are selected, respectively. To improve the statistics, those detectors of GECAM and GBM with incident angle of SGR J1935+2154 less than 60 degrees are selected, and the combined light curves from these detectors are used.
The time accuracy of the GBM and GECAM is within 10 $\mu$s \citep{meegan2009fermi,xiao2022ground}, thus the time system does not need to be corrected.

In this work, 55, 100 and 581 bursts from SGR J1935+2154 are collected for HXMT, GECAM, and GBM, respectively. Note that, since GECAM was launched in December 2020, its bursts are collected after this date. Besides, the samples for HXMT we selected are only from April 28 2020 to May 24 2020 during the month-long dedicated pointing observation \citep{cai2022insight}.

\subsection{Calculation of spectral lags}
The improved Li-CCF method (i.e. MCCF) (\citealp{li2004timescale}; \citealp{2021ApJ...920...43X}) is defined as
\begin{equation}
\begin{split}\label{equ:MCCF}
{\rm MCCF}(k,\Delta t)=\frac{1}{M_{\Delta t}} \sum_{m=1}^{M_{\Delta t}} \sum_{i} u_m(i;\Delta t)\upsilon_{m+k}(i;\Delta t)/\sigma_ u \sigma_\upsilon.
\end{split}
\end{equation}

Based on the high time resolution initial lightcurves of ($x(j;\delta t)$ and $y(j;\delta t)$), the phase factor m = 1, 2, $\cdots$, $M_{\Delta t}$, and $u_m(\Delta t)$ and $\upsilon_m(\Delta t)$ are the background-subtracted series of $x_m(\Delta t)$ and $y_m(\Delta t)$ by re-binning the initial series, then obtain the lightcurves with a optimized timebin $\Delta t=M_{\Delta t} \delta t$ (from 1 ms to 20 ms in this work),
\begin{equation}
\begin{split}\label{equ:improved MCCF}
u_m(i;\Delta t)=x_m(i;\Delta t)-b_{x_m}(i;\Delta t),\\
\upsilon_m(i;\Delta t)=y_m(i;\Delta t)-b_{y_m}(i;\Delta t),\\
\sigma^{2}_u=\sum_{i} {u_m(i;\Delta t)}^{2},\\ \sigma^{2}_\upsilon=\sum_{i} {\upsilon_m(i;\Delta t)}^{2},
\end{split}
\end{equation}
where $b_{x_m}$ and $b_{y_m}$ are the background counts of lightcurves $x_m(\Delta t)$ and $y_{m}(\Delta t)$, respectively. The spectral lag between the two lightcurves $y_{m} (i;\Delta t)$ and $x_{m}(i;\Delta t)$ on $\Delta t$ is $\tau(\Delta t)={k_{\rm max} \delta t}$.

Since the spectral lag is defined between the `same' pulse in different energies, this means that the maximum spectral lag usually should not exceed the pulse half-width. The duration of the burst is typically $\lesssim$ 100 ms, thus in this work we set the maximum time delay to -60 ms to 60 ms to improve the statistical error for the weak bursts. All spectral lags from the relatively bright bursts are only on the order of milliseconds, meaning that this range setting is reasonable. It is worth noting that this range does not affect the conclusions of our work, because the bursts with large errors will be filtered out when doing statistical analysis.

$\delta t$ is set to 0.1 ms in this work, considering both the accuracy and time consumption of calculation. To estimate the uncertainty of spectral lag, we perform a Monte Carlo (MC) simulation for the observed initial lightcurves based on Poisson probability distribution. The standard deviation of the distribution can be used as 1$\sigma$ level because the distribution approximately follow a Gaussian distribution in most cases \citep{2021ApJ...920...43X}.

\section{Results}

\subsection{The spectral lag between 10-20 keV and 60-100 keV}\label{section_lag2}
Since the photon energy from SGR J1935+2154 is mainly below 100 keV, we first investigate the spectral lag between the lowest and highest energies. Table \ref{all_lags} shows the spectral lags between 10-20 keV and 60-100 keV for the HXMT, GECAM and GBM, respectively.
Note that spectral lags with errors $>$ 20 ms are not well constrained because the maximum range of spectral lag is set to 60 ms.

The left panel of Fig.~\ref{lag_2} shows the distribution of the spectral lags (with the error$<$20 ms) between 10-20 keV and 60-100 keV for each burst observed by {\it Insight}-HXMT or GECAM or Fermi/GBM. The size of burst samples for GBM, GECAM and HXMT are 358, 32 and 48, respectively, noting that most of the samples from the three satellites are different due to their different orbital positions (i.e. different fields of view) and sensitivities. (e.g. the burst sample of HXMT has much weaker bursts than that of GBM and GECAM). However, it can be found that their distribution ranges are generally consistent. The GBM burst sample is larger and its distribution concentrates around the peak of $\sim$ 1.3 ms.
The percentage of bursts with lag $<$ 0 (i.e. low-energy photons arrive earlier) at significance $>$1 $\sigma$ is 7\% for GBM, 3\% for GECAM and 13\% for HXMT.

To test whether the distributions are dependent on the sample selection, we also show the distributions of spectral lags with error $<$ 5 ms in the right panel of Fig.~\ref{lag_2}. The sizes of these samples for GBM, GECAM and HXMT are 197, 10 and 35, respectively. The peak is the same as above at 1.3 ms.
The percentage of lag $<$ 0 at significance $>$1 $\sigma$ is 5\% for GBM, 10\% for GECAM and 14\% for HXMT. 

Besides, the distribution of spectral lags for the sample observed by GBM approximately follows a Gaussian function, with mean and standard deviation of 1.3 ms and 2.3 ms for the bursts with spectral lag error $<$ 20 ms, 1.4 ms and 1.7 ms for the samples with error $<$ 5 ms. Thus, it seems that the distributions are consistent for samples with different errors, and the samples with smaller errors have a narrower distribution.

We also investigate the significance of the non-zero spectral lag (i.e. the center spectral lag value/the error). For the same burst with multiple satellites jointly observed, we choose the result with the smallest error, the left panel in Fig.~\ref{sigmas} shows the distribution of the significance of non-zero spectral lags between 10-20 keV and 60-100 keV for the bursts with error$<$10 ms, where the percentages of significance > 1, 2 and 3 $\sigma$ are about 46\% (138/301), 19\% (57/301) and 8\% (25/301), respectively. 

\subsection{Spectral lag fitting for each burst}\label{fit_each}
The dependence of spectral lag with energy is also valuable for better understanding of the radiation mechanism (e.g. \citealp{zhang2009discerning}, \citealp{shao2017new}). The 1-6 keV, 10-20 keV and 8-15 keV are selected for HXMT, GECAM and GBM as the lowest energy bands respectively. Higher energy lightcurves are 6-10, 10-15, 15-30, 30-35, 35-45, 45-60, 60-100 keV for HXMT, 20-25, 25-30, 30-35, 35-45, 45-60, 60-100 keV for GECAM, and 15-20, 20-25, 25-30, 30-35, 35-45, 45-60, 60-100 keV for GBM, and then the spectral lag between each higher energy and the reference lowest energy band are calculated. We then used the $optimize.curve\_fit$ package for Python to fit a linear function to the spectral lag vs. energy for each burst. The uncertainties on the fit parameters are obtained from the diagonal of the covariance matrix, multiplied by a factor of $\sqrt{\chi^2/dof}$ in analogy with Ukwatta et al. \citep{ukwatta2012lag}.

We find that the energy-spectral lag relations for most bursts can be fitted well with $t_{\rm lag}(E) =\alpha (E/{\rm keV})+C$. Fig. A2 - A8 in Supplementary material (available online) show the fitting plots with the error of $\alpha<0.05$ by Fermi/GBM, GECAM and HXMT. The complete fitting results are listed in Table A2 in Supplementary material.

Panel (a) of Fig.~\ref{alphas} shows the distribution of the slope $\alpha$ (with error $<$0.05) for each burst observed by {\it Insight}-HXMT or GECAM or Fermi/GBM. 
Note that most of the samples from the three satellites are different. It can be found that their distribution ranges are also consistent. The distribution of $\alpha$ for the GBM sample follows one Gaussian distribution with mean and standard deviation of 0.02 and 0.04 ms/keV respectively, but there may exist three peaks at $\sim$ -0.009, 0.013 and 0.039 ms/keV.
Besides, the distribution of $\alpha$ for GECAM shows consistent peaks with GBM. The distribution of $\alpha$ for HXMT is mainly concentrated around the first peak of -0.009, though the statistics is low. $\alpha<0$ means that the lower energy photons arrive earlier (note that there is an overall effect here), and the percentage of $\alpha<$ 0 at significance $>$1 $\sigma$ is 9\% for GBM, 13\% for GECAM, and 11\% for HXMT, respectively.

We also test whether the distributions are dependent on the sample selection. Panel (b) of Fig.~\ref{alphas} shows the distribution of the slopes with the error of $\alpha$ $<$0.02.
The mean and standard deviation of the Gaussian of fitting for GBM are 0.02 and 0.02 ms/keV, respectively, which are consistent with the results of the samples with the error of $\alpha$ $<$0.05 but more narrow.
We also find that three Gaussians (with peaks at $\sim$ -0.009, 0.013 and 0.039 ms/keV) can fit this distribution better than one Gaussian (see residuals in panels (c) and (d) of Fig.~\ref{alphas}); the fitted parameters are listed in Table A1 and the errors are obtained by a bootstrap method. The significance of these three Gaussians are 1.9$\sigma$, 5.1$\sigma$ and 3.7$\sigma$, respectively.

It is worth noting that with the mean $\alpha$ = 0.02 ms/keV, the spectral lag between 10-20 and 60-100 keV should be about $t_{\rm lag}=\alpha (E/{\rm keV})=0.02*((100+60)/2-(20+10)/2)\ {\rm ms}=1.3$ ms, which is perfectly in agreement with the result obtainted in Section \ref{section_lag2} (see Fig.~\ref{lag_2}). This also verifies the linear energy-dependence of spectral lag.

The right panel in Fig.~\ref{sigmas} shows the distribution of the significance of non-zero slopes of the energy-spectral lag relation $t_{\rm lag}(E) =\alpha (E/{\rm keV})+C$ for the bursts with the spectral lag error$<$10 ms, where the percentages of significance > 1, 2 and 3 $\sigma$ are about 61\% (183/301), 30\% (89/301) and 18\% (53/301), respectively. Besides, we also assessed if the linear model is statistically required with respect to the constant model with an F-test \citep{rawlings1998applied}, the percentages of significance of the  non-constant linear model > 1, 2 and 3 $\sigma$ are about 76\% (229/301), 41\% (124/301) and 15\% (44/301), respectively. It can be found that the results obtained by the two different methods are general consistent.

\subsection{Joint fit of the spectral lag for bursts}
Since the distribution of $\alpha$ for GBM roughly follows a Gaussian distribution, it is reasonable to suppose that the spectral lags of majority bursts are essentially the same, but only the statistical uncertainties caused the differences. Therefore, we investigate the average spectral lags in the different energy bands of the bursts from SGR J1935+2154. We fit the spectral lags of multiple bursts in each higher energy band compared to the lowest energy band, respectively (see Fig. A1). To improve the statistics, bursts with spectral lag errors of $<$ 20 ms between 10-20 and 60-100 keV are selected (see Table A2 and left panel of Fig.~\ref{lag_2}), resulting in a total of 358 bursts. Then, we fit the energy-spectral lag of joint fit using the $t_{\rm lag}(E) =\alpha (E/{\rm keV})+C$, and the result is $t_{\rm lag}(E)=0.019^{+0.001}_{-0.001}(E/{\rm keV})-0.228$ ms (see the left panel of Fig.~\ref{lag_fit_joint}), which is consistent with the distribution of the spectral lag obtained by fitting individual bursts in Section \ref{fit_each}.

To understand the linear relationship of $t_{\rm lag}(E) =\alpha (E/{\rm keV})+C$, we make a simple simulation by assuming that the spectral lag is caused by spectral evolution during a burst. Since the most bursts from SGR J1935+2154 can be fitted with a single blackbody or a double blackbody \citep{lin2020fermi,gougucs2020persistent,kaneko2021fermi,2022ApJS..260...25C}. Although some bright bursts from SGR J1935+2154 in the Burst Forest \citep{kaneko2021fermi} are best fit to a double blackbody model and the dimmer to a single BB model, we consider a single blackbody instead of a double blackbody in the simulation to avoid the degeneracy due to the limited statistics of the linear relationship. Besides, the $kT$ of blackbody depends on the temperature of the blackbody-emitting plasma (e.g. \citealp{dai2020magnetar, lin2020fermi}).
Assuming that the energy spectrum of a burst is a single blackbody at each small time interval and the temperature varies with time linearly, the shape of the light curve is the fast\-rising exponential decay (FRED) \citep{norris1996attributes}, the background is smooth, and the uncertainty follows a Poisson distribution. We generate event-by-event simulation data including arrive time and energy information by reject sampling, the arrival time is generated using the $pyipn$ package designed by Burgess et al. \citep{burgess2021nazgul}. We simplified the simulation step by not considering the energy response of the instrument, that is, assuming that the energy of the incident photon is what we observed. Then the spectral lag is calculated between different energy ranges, and we find that the spectral lag also varies linearly with energy. Specifically, a temperature decrease rate of $kT=-12\ (t/{\rm s})+14$ keV (note the rate of temperature variation is dependent on the initial temperature) results in a spectral lag increase rate of $\sim$ 0.02 ms/keV (see the right panel of Fig.~\ref{lag_fit_joint}), which suggests the existence of a linear change in temperature with time for most bursts. Similarly, if the temperature rises linearly with time, it will also lead to a spectral lag with negative linear dependence on energy.

\begin{figure}
\centering
\includegraphics[width=\columnwidth]{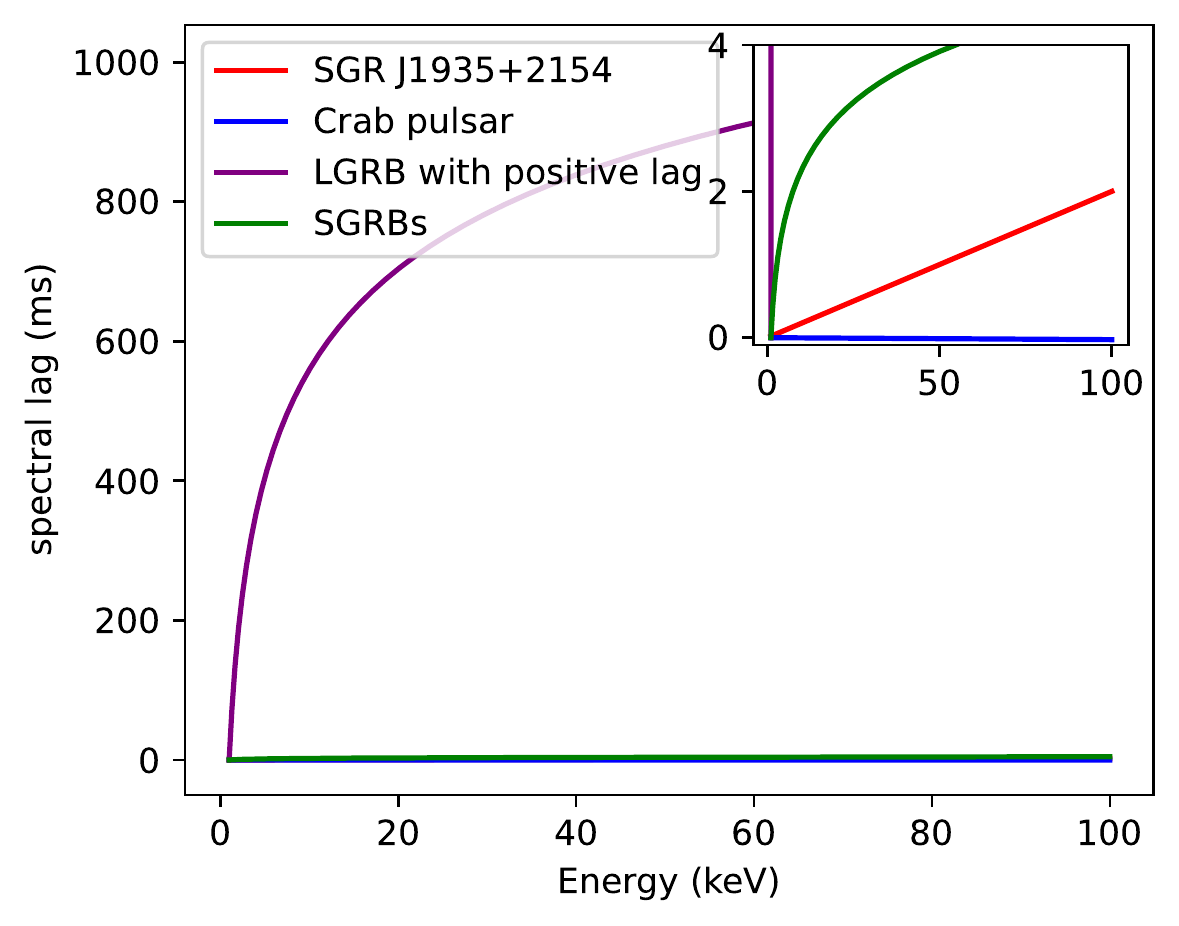}
\caption{The comparison of spectral lag-energy relation of bursts from SGR J1935+2154 ($t_{\rm lag}=2\times 10^{-2} (E/{\rm keV})$ ms), Crab pulsar ($t_{\rm lag}=-2.7\times 10^{-4} (E/{\rm keV})$ ms), LGRB ($t_{\rm lag}=2.7\times 10^3 (1-(E/{\rm keV})^{-0.1})$ ms) and SGRB ($t_{\rm lag}=1\times \ln (E/{\rm keV})$ ms). Note that the LGRB is given as an example of a positive spectral lag, which may be different for different LGRBs. }\label{dif_lag}
\end{figure}

\section{Discussion and conclusion}
We present for the first time a detailed analysis of the spectral lag of a large sample of bursts from SGR J1935+2154 observed by {\it Insight}-HXMT, GECAM and Fermi/GBM from July 2014 to Jan 2022 using the improved Li-CCF method, which can make use of more temporal information contained in high time resolution data and obtain more accurate values in milliseconds. We discover that the spectral lags of most bursts (about 61\% with non-zero significance >1$\sigma$) from SGR J1935+2154 are linearly dependent on the photon energy ($E$) with $t_{\rm lag}(E)=\alpha (E/{\rm keV})+C$. The percentage of $\alpha<$ 0 at significance $>$1 $\sigma$ is 9\% for GBM, 13\% for GECAM and 11\% for HXMT. The distribution of $\alpha$ approximately follows a Gaussian distribution with mean and standard deviation of 0.02 ms/keV and 0.02 ms/keV, respectively. Besides, the 
mean spectral lag of the bursts from SGR J1935+2154 between 10-20 and 60-100 keV is $\sim$ 1.3 ms, which is consistent with the result of $t_{\rm lag}=0.02(E/{\rm keV})+C$ ms. This also verifies the existence of linear energy-dependence of spectral lag.  

We also find that the distribution of $\alpha$ can be better fitted with three Gaussians with mean values (or peaks) of $\sim$ -0.009, 0.013 and 0.039 ms/keV. Because the energy responses of different instruments are different and the burst samples of the three satellites are also different, it is difficult to be completely consistent between different detectors in sub-millisecond or milliseconds spectral lag measurements, however, these three peaks are consistent in the distributions for GBM and GECAM, thanks to their similar energy responses. But the first peak is more obvious in the distribution of HXMT, which is composed of much weaker bursts. We tried to compare the duration and energy spectra of those bursts around the three peaks, but no significant differences were found thus far; further analysis is in progress (Xiao et al., in preparation).

 It is worth noting that this relationship $t_{\rm lag}(E)=\alpha (E/{\rm keV})+C$ may not be true for all bursts from SGR J1935+2154. The first reason is that it is difficult to measure the spectral lags for many weak bursts, which are therefore not studied in detail in this work. The second is that some bursts do not follow the relation well, but it is difficult to judge whether it is caused by statistical errors. Finally, the spectral lag calculated by the cross-correlation method is the `average effect' over the entire duration of the burst, which implies that the spectral lag may be dominated by a small fraction of the lightcurve with higher counts. Therefore, the $t_{\rm lag}(E)=\alpha (E/{\rm keV})+C$ relation may only exist in some portion of the lightcurve in a burst, or there may exist both increase ($\alpha>0$) or decrease ($\alpha<0$) relations within a burst, which could also be the reason why some bursts do not follow the relation well. For example, for the burst at UTC 2021-09-12T00:34:37.414, we fit the time resolved spectrum using single Blackbody model, the Blackbody temperature decreases linearly with time (see Fig.~\ref{spec_fit}), thus the spectral lag is linearly increasing with energy ($\alpha>0$). But for the burst at UTC 2022-01-15T17:21:59.297, the Blackbody temperature rises and then falls with time, but since the rise is more dominant, the $\alpha<0$ (i.e. the the lower energy photons arrive earlier).

We investigated the consistency of the spectral lag measurements for bursts observed by GECAM and GBM, including four bursts with errors less than 5 ms (2021-02-16T22:20:39.600,2021-07-07T00:33:31.640,2021-09-12T00:34:37.232,2022-01-15T17:21:59.283), all of which are consistent within the error range. In this work, we did not consider the effect of the energy resolutions \citep{xiao2022robust} of the instruments, which are about 12\% @59.5 keV \citep{li2021inflight} for GECAM, 19\% @59.5 keV for HXMT/HE \citep{zhang2020overview} and 15\% @100 keV for GBM \citep{meegan2009fermi}. However, after considering the 7 keV energy resolution for Swift/BAT as in \citealp{xiao2022robust}, the spectral lag varies by about only 10\%, thus this effect does not significantly affect our conclusions. Besides, since we mainly study the relationship between spectral lag and energy, the effect should be much smaller.

We also investigated whether the linear energy-dependent spectral lag is a unique phenomenon for the SGR J1935+2154, we randomly selected two relatively bright bursts (UTC 2009-01-22T01:03:53.93 and 2009-01-22T01:14:47) observed by GBM from another magnetar, SGR J1550-5418, and find that there are also linear energy-dependent spectral lags using same method. A more detailed analysis for bursts from other magnetars is beyond the scope of this work, but it at least shows that linear energy-dependent spectral lag is not uncommon in magnetars. On the other hand, we compare the spectral lag-energy relation of the bursts from SGR J1935+2154 ($t_{\rm lag}(E)=2\times 10^{-2}(E/{\rm keV})$ ms), Crab pulsar ($t_{\rm lag}(E)=-2.7\times 10^{-4}(E/{\rm keV})$ ms), LGRB ($t_{\rm lag}(E)=2.7\times 10^{3}(1-(E/{\rm keV})^{-0.1})$ ms) and SGRB ($t_{\rm lag}(E)=1\times \ln (E/{\rm keV})$ ms) in Fig.~\ref{dif_lag}. 
We find that the the absolute value of the slope $\alpha$ of SGR J1935+2154 is about one hundred times more than that of Crab, but their signs are opposite. Interestingly, the magnetic field of SGR J1935+2154 (i.e. $\sim 10^{14}$ G \citep{israel2016discovery}) is also almost one hundred times more than that of Crab pulsar (i.e. $\sim 10^{12}$ G \citep{buhler2014surprising}), and the period of SGR J1935+2154 (i.e. 3.2 s \citep{israel2016discovery}) is also about one hundred times more than that of Crab pulsar (i.e. $\sim$ 34 ms \citep{lyne199323}). However, it is not clear whether this is just a coincidence or there are some physical implications. On the other hand, the relationship between SGR J1935+2154 and SGRBs (i.e. $t_{\rm lag}(E)=A \ln (E/{\rm keV})+C$) is very different. The latter may come from the jet `curvature' effect \citep{zhang2009discerning}, while the former may be from a fireball (e.g. \citealp{dai2020magnetar,yang2021fast}). However, the values of their spectral lags are comparable, that is $\sim$ 3.2 ms in SGRBs between 10-20 and 60-100 keV, which can be obtained by $t_{\rm lag}(E)=A \ln (E/{\rm keV})+C$ relation \citep{minaev2014catalog,xiao2022robust}.

Finally, we notice that the bursts from SGR J1935+2154 are more suitable for spectral lag measurements and studies than GRBs, because there may be different spectral lags in different pulses across GRBs \citep{hakkila2008correlations}, while the bursts from SGR J1935+2154 usually have a single pulse, which makes the measurement of spectral lag more robust. Therefore, the measurement of spectral lag should be an important way in the timing analysis for SGR-like GRBs, from which the energy spectrum evolution could be inferred. Especially for some bursts like those from SGR J1935+2154 with bright but short duration (e.g. 0.1 s), the time resolved spectrum is different due to a few net counts, the spectral lag can be an indicator. Therefore, spectral lag analysis could become an important tool for understanding the radiation mechanism of SGRs.

\section*{Acknowledgments}
We thank the anonymous reviewer for a careful reading of our manuscript and insightful comments and suggestions, especially regarding significance estimates and study implications. This work is supported by the National Key R\&D Program of China (2022YFF0711404, 2021YFA0718500), the Scientific Research Project of the Guizhou Provincial Education (Nos. KY[2022]123, KY[2022]132, KY[2022]137), and partially supported by International Partnership Program of Chinese Academy of Sciences (Grant No. 113111KYSB20190020). 
The authors also thank supports from 
the Strategic Priority Research Program on Space Science, the Chinese Academy of Sciences (Grant No.
XDA15010100, 
XDA15360100, XDA15360102, 
XDA15360300, 
XDA15052700), 
 the National Natural Science Foundation of China (Projects: 12061131007 
 and Grant No. 12173038), the Foundation of Education Bureau of Guizhou Province, China (Grant No. KY (2020) 003), Guizhou Provincial Science and Technology Foundation (Nos. ZK[2022]304), the Major Science and Technology Program of Xinjiang Uygur Autonomous Region (No.2022A03013-4).
 S. Xiao is grateful to W. Xiao, G. Q. Wang and J. H. Li for their useful comments. 

\newpage
\section*{DATA AVAILABILITY}
The data underlying this article will be shared on reasonable request
to the corresponding author.
\bibliographystyle{mnras}
\bibliography{ref0} 
\newpage
\section*{}

\newpage
\renewcommand\thefigure{\Alph{section}A\arabic{figure}}

\setcounter{figure}{0}

\begin{figure*}
\centering
\includegraphics[width=1.2\columnwidth]{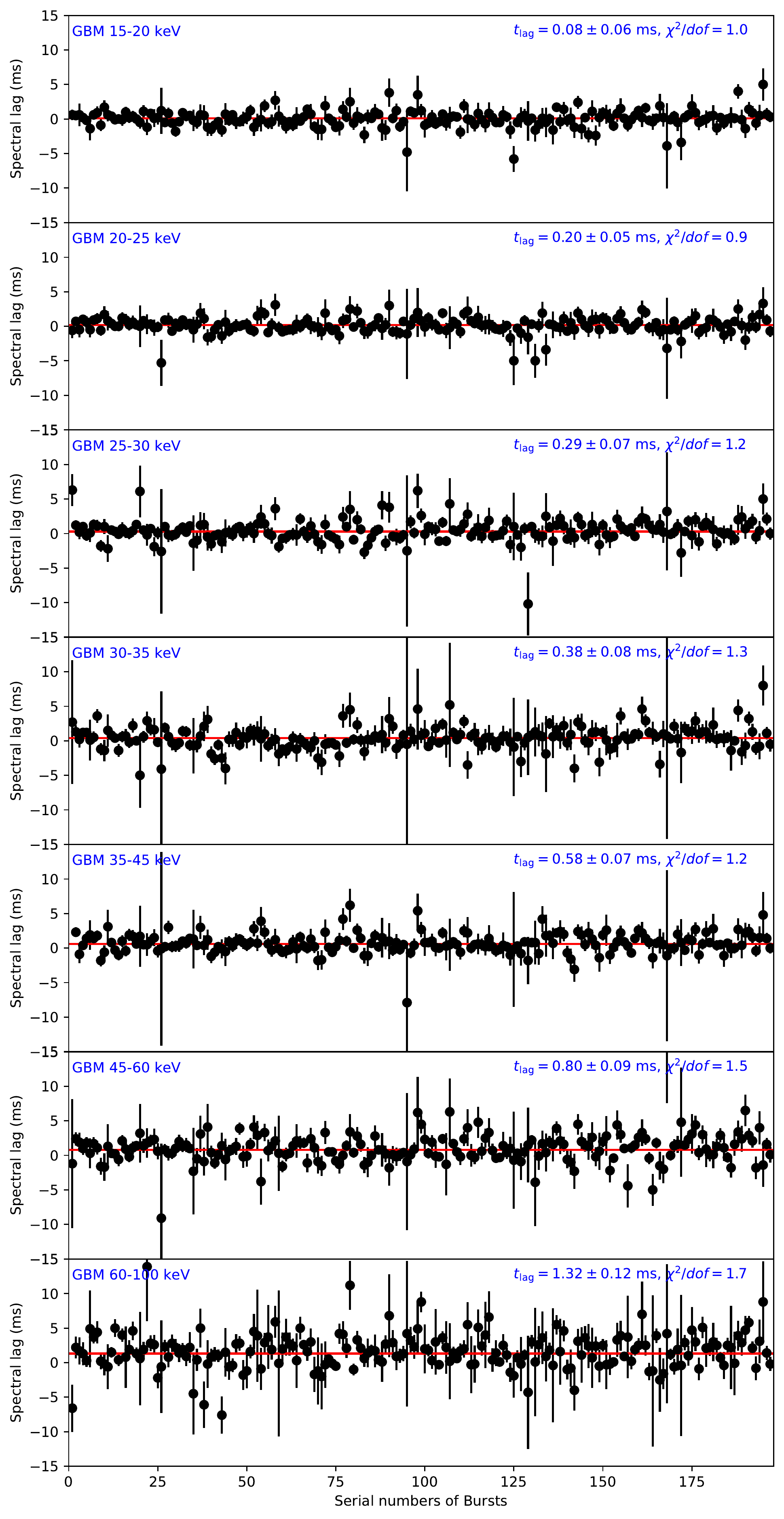}
\caption{The spectral lags between different energy bands compared to lowest energy band of 197 bursts observed by Fermi/GBM are fitted, respectively. Bursts with errors of spectral lag $<$ 20 ms between 10-20 and 60-100 keV are selected.}\label{joint_fit}
\end{figure*}

\begin{figure*}
\centering
\begin{minipage}[t]{0.96\textwidth}
\centering
\includegraphics[width=\columnwidth]{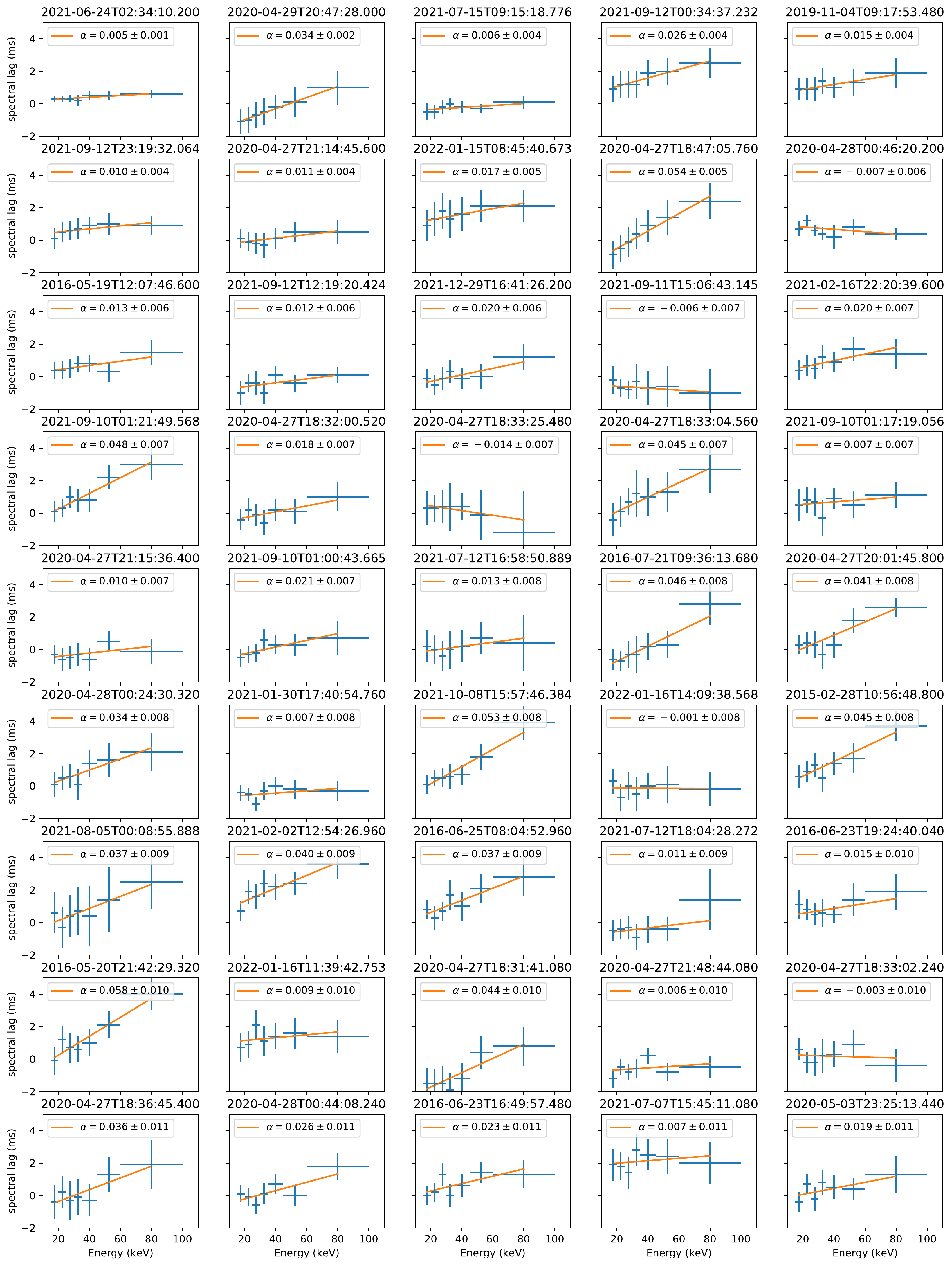}
\end{minipage}
\caption{Energy-spectral lag relationship observed by Fermi/GBM and the results of fitting with $t_{\rm lag}(E) =\alpha (E/{\rm keV})+C$ with the error of $\alpha<0.05$.}\label{lag_each0}
\end{figure*}

\begin{figure*}
\centering
\begin{minipage}[t]{0.96\textwidth}
\centering
\includegraphics[width=\columnwidth]{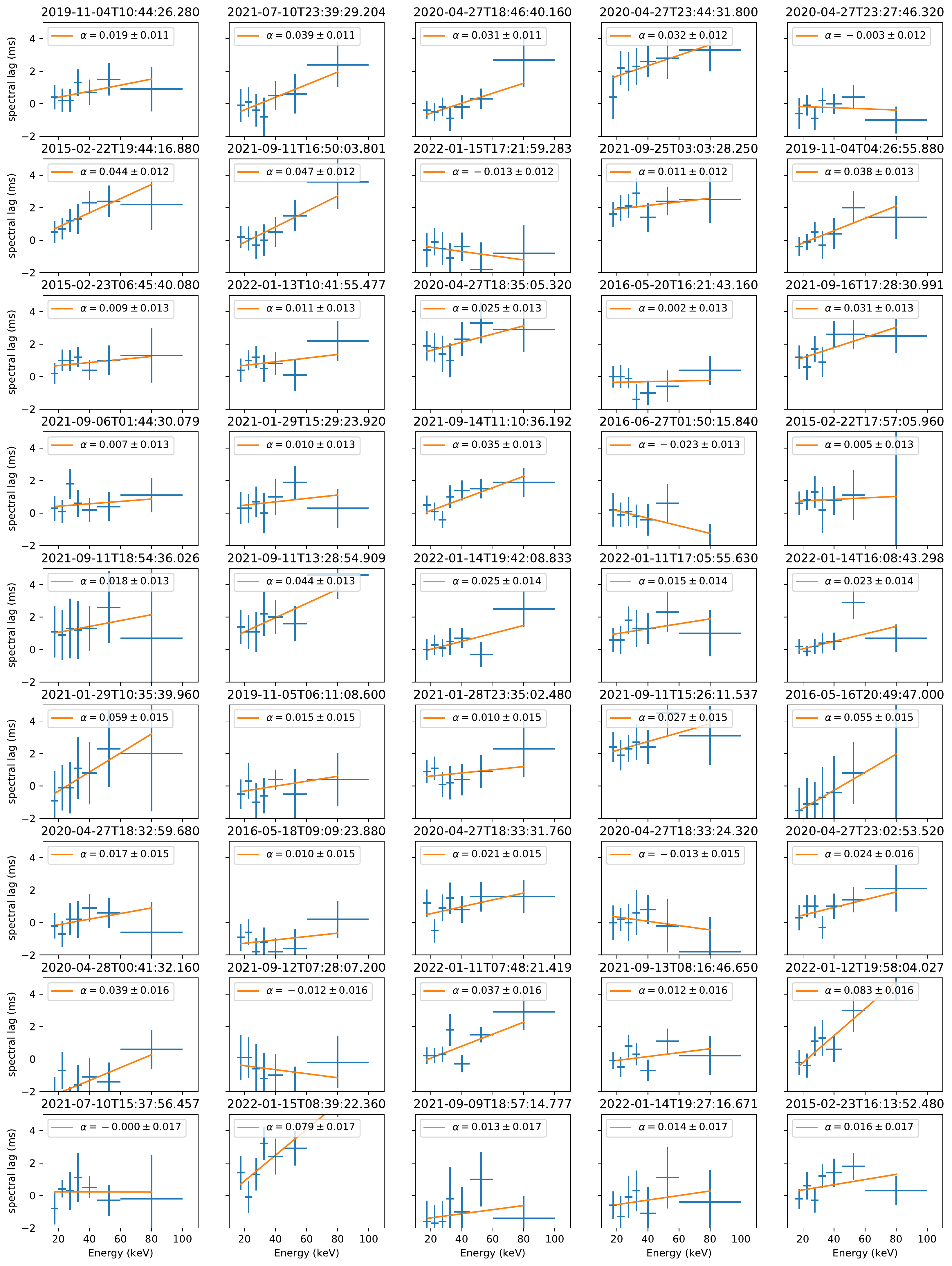}
\end{minipage}
\caption{Continuation of Fig.~\ref{lag_each0}.}\label{lag_each1}
\end{figure*}

\begin{figure*}
\centering
\begin{minipage}[t]{0.96\textwidth}
\centering
\includegraphics[width=\columnwidth]{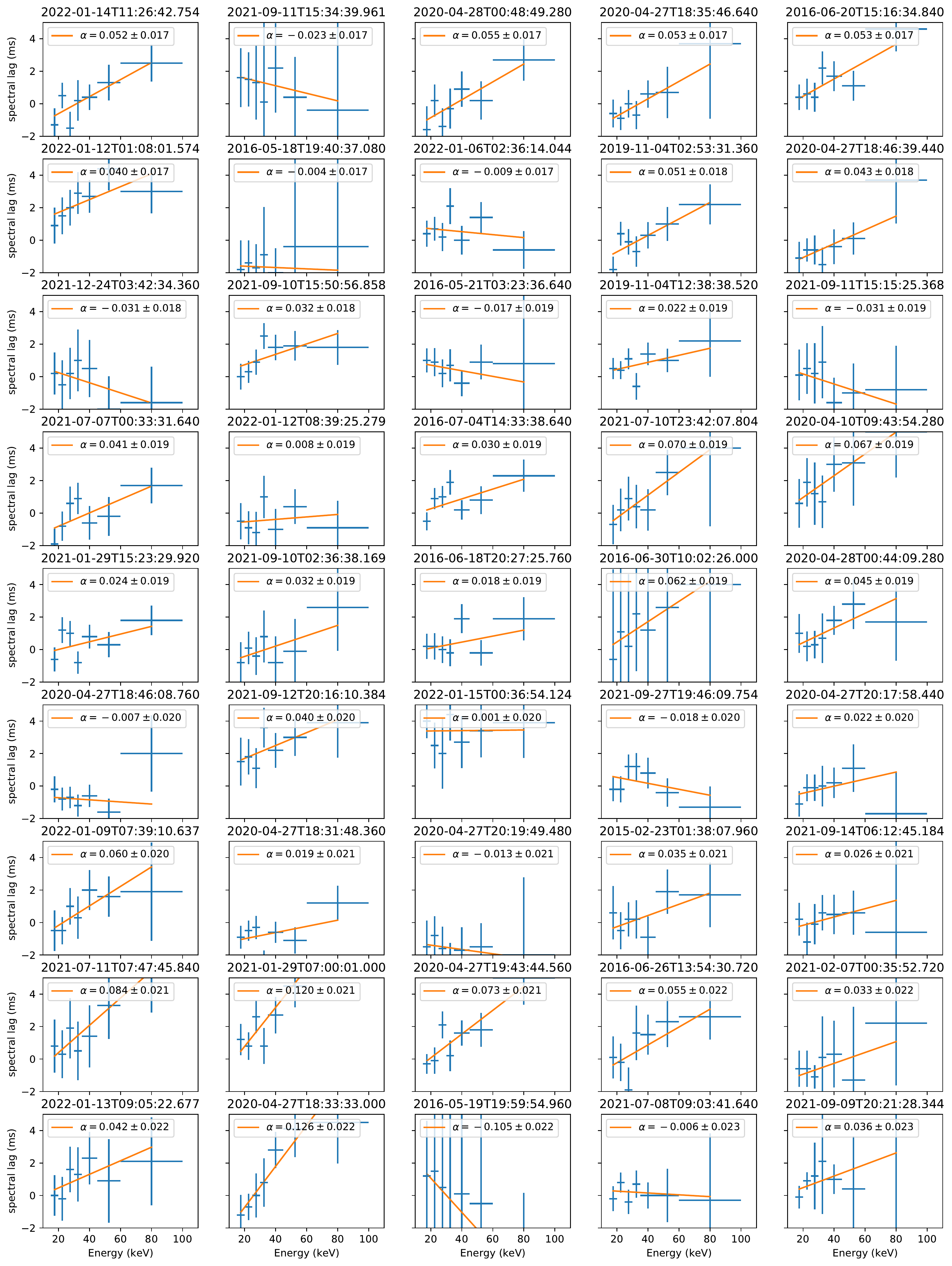}
\end{minipage}
\caption{Continuation of Fig.~\ref{lag_each0}.}\label{lag_each2}
\end{figure*}

\begin{figure*}
\centering
\begin{minipage}[t]{0.96\textwidth}
\centering
\includegraphics[width=\columnwidth]{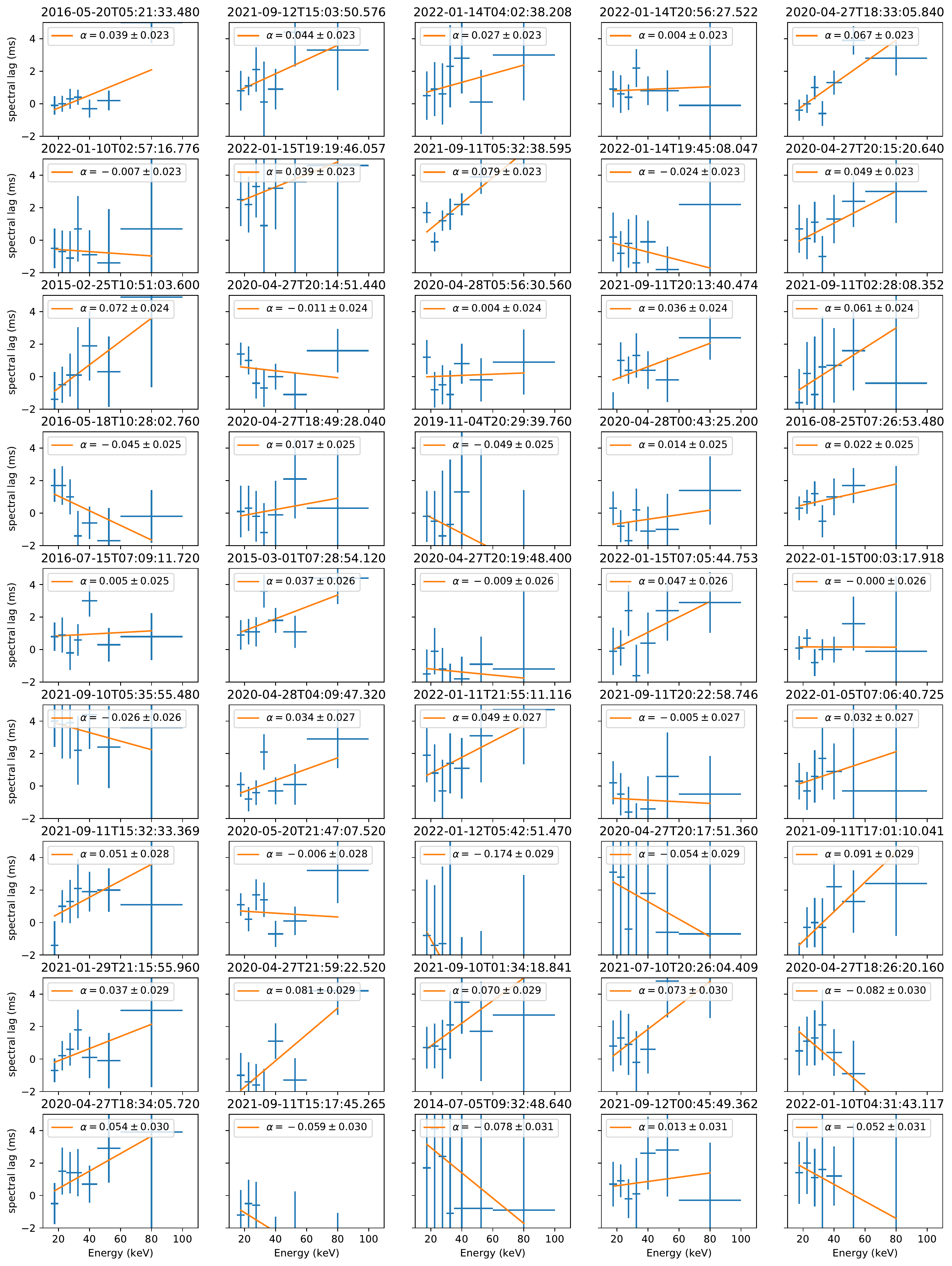}
\end{minipage}
\caption{Continuation of Fig.~\ref{lag_each0}.}\label{lag_each1}
\end{figure*}

\begin{figure*}
\centering
\begin{minipage}[t]{0.96\textwidth}
\centering
\includegraphics[width=\columnwidth]{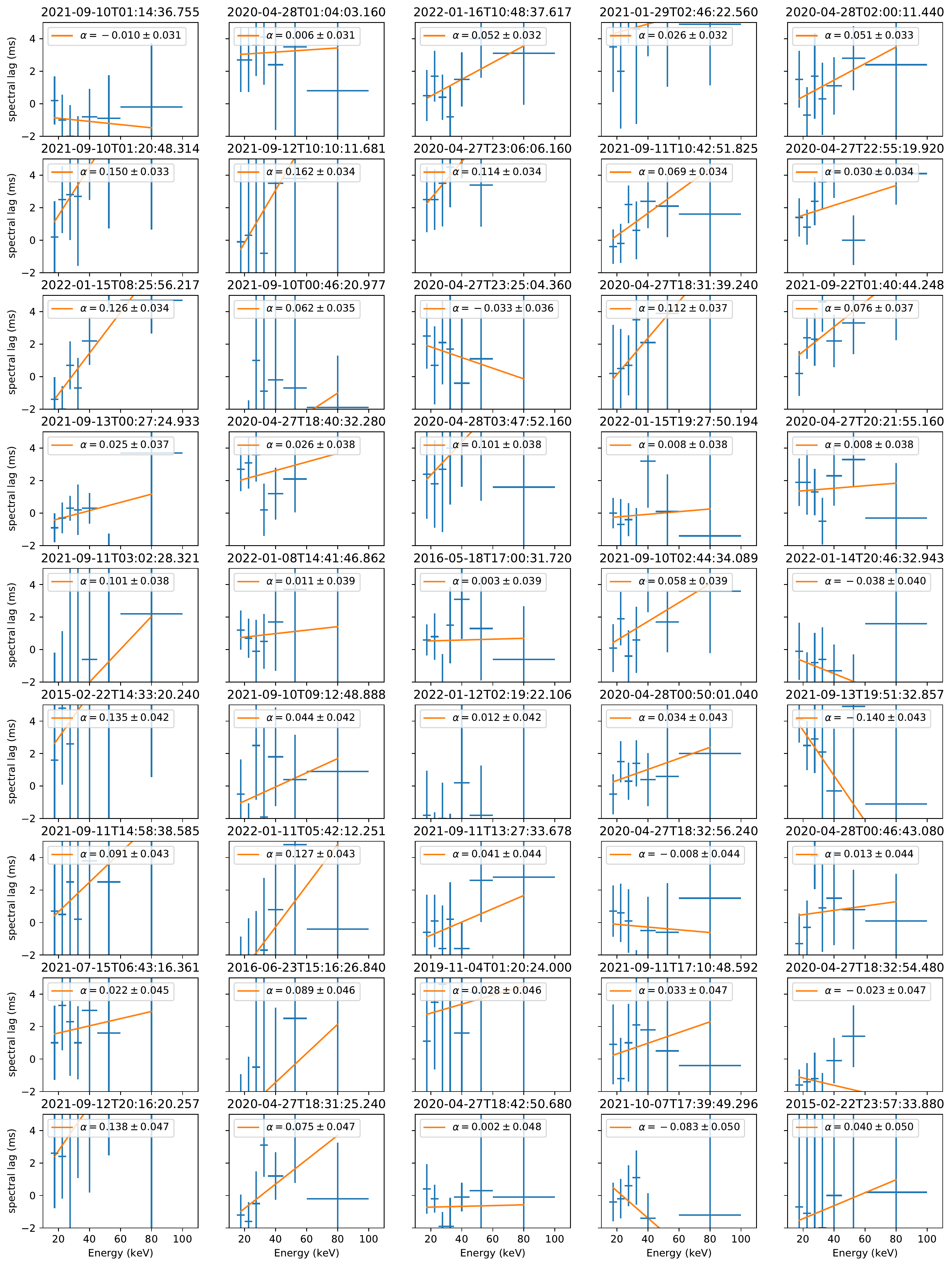}
\end{minipage}
\caption{Continuation of Fig.~\ref{lag_each0}.}\label{lag_each5}
\end{figure*}

\newpage
\begin{figure*}
\centering
\begin{minipage}[t]{0.96\textwidth}
\centering
\includegraphics[width=\columnwidth]{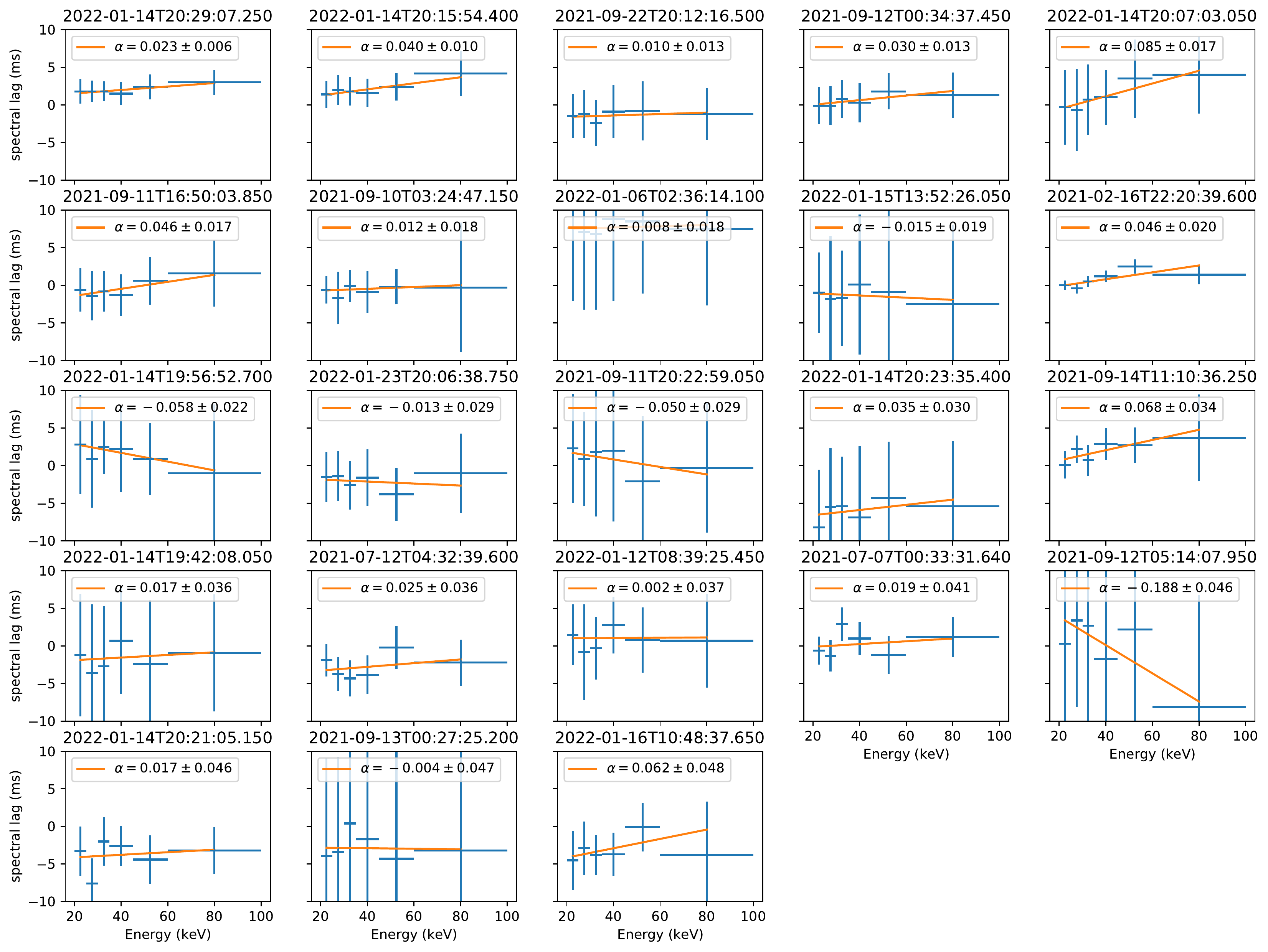}
\end{minipage}
\caption{Energy-spectral lag relationship observed by GECAM and the results of fitting with $t_{\rm lag}(E) =\alpha (E/{\rm keV})+C$ with the error of $\alpha<0.05$.}\label{lag_each_gecam}
\end{figure*}

\newpage

\begin{figure*}
\centering
\begin{minipage}[t]{0.96\textwidth}
\centering
\includegraphics[width=\columnwidth]{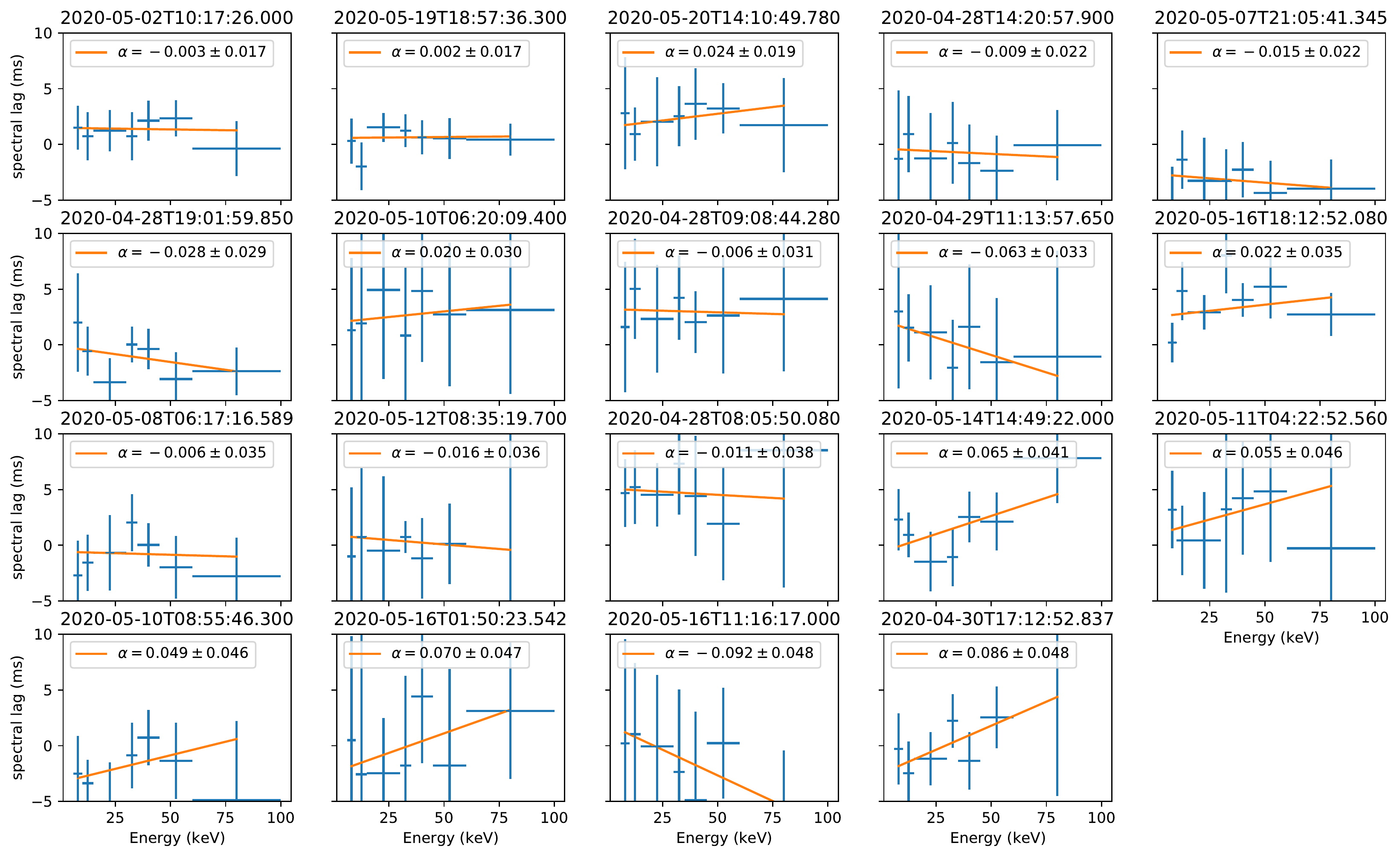}
\end{minipage}
\caption{Energy-spectral lag relationship observed by {\it Insight}-HXMT and the results of fitting with $t_{\rm lag}(E) =\alpha (E/{\rm keV})+C$ with the error of $\alpha<0.05$.}\label{lag_each_hxmt}
\end{figure*}
\newpage

\newpage

\renewcommand\thetable{\Alph{section}A\arabic{table}}

\onecolumn
\begin{longtable}{ccccccc}
\endfirsthead
\hline
\endhead
\hline
\endfoot
\hline
\endlastfoot\\
\caption{Spectral lags between 10-20 keV and 60-100 keV of the bursts from SGR J1935+2154 observed by {\it Insight}-HXMT ($t_{\rm lag({\rm HXMT})}$), GECAM ($t_{\rm lag({\rm GECAM})}$) and Fermi/GBM ($t_{\rm lag({\rm GBM})}$), as well as fitting results ($\alpha_{\rm HXMT}$, $\alpha_{\rm GECAM}$ and $\alpha_{\rm GBM}$) using $t_{\rm lag}(E) =\alpha (E/{\rm keV})+C$.}\label{all_lags}\\
\hline
Trigger time (UTC)& $t_{\rm lag({\rm HXMT})}$ (ms) & $t_{\rm lag({\rm GECAM})}$& $t_{\rm lag({\rm GBM})}$& $\alpha_{\rm HXMT} ({\rm ms/keV})$ &$\alpha_{\rm GECAM}$ &$\alpha_{\rm GBM}$ \\

\hline
2014-07-05T09:32:48.640 & -              & -              & $-2.4\pm13.1$  & -                & -                & $-0.078\pm0.031$ \\
2014-07-05T09:37:34.480 & -              & -              & $-0.2\pm27.1$  & -                & -                & $0.203\pm0.164$  \\
2014-12-27T02:16:21.280 & -              & -              & $-6.8\pm4.3$   & -                & -                & $-0.078\pm0.096$ \\
2015-02-22T13:33:00.000 & -              & -              & $-32.4\pm28.3$ & -                & -                & $0.133\pm0.408$  \\
2015-02-22T14:33:20.240 & -              & -              & $9.2\pm7.2$    & -                & -                & $0.135\pm0.042$  \\
\hline

\end{longtable}


\bsp	
\label{lastpage}
\end{document}